\documentclass{iopart}
\usepackage{graphicx,color}
\usepackage{iopams}

\newcommand{\figsch}{
\begin{figure}[htbp]
\begin{center}
	\includegraphics[width=6cm]{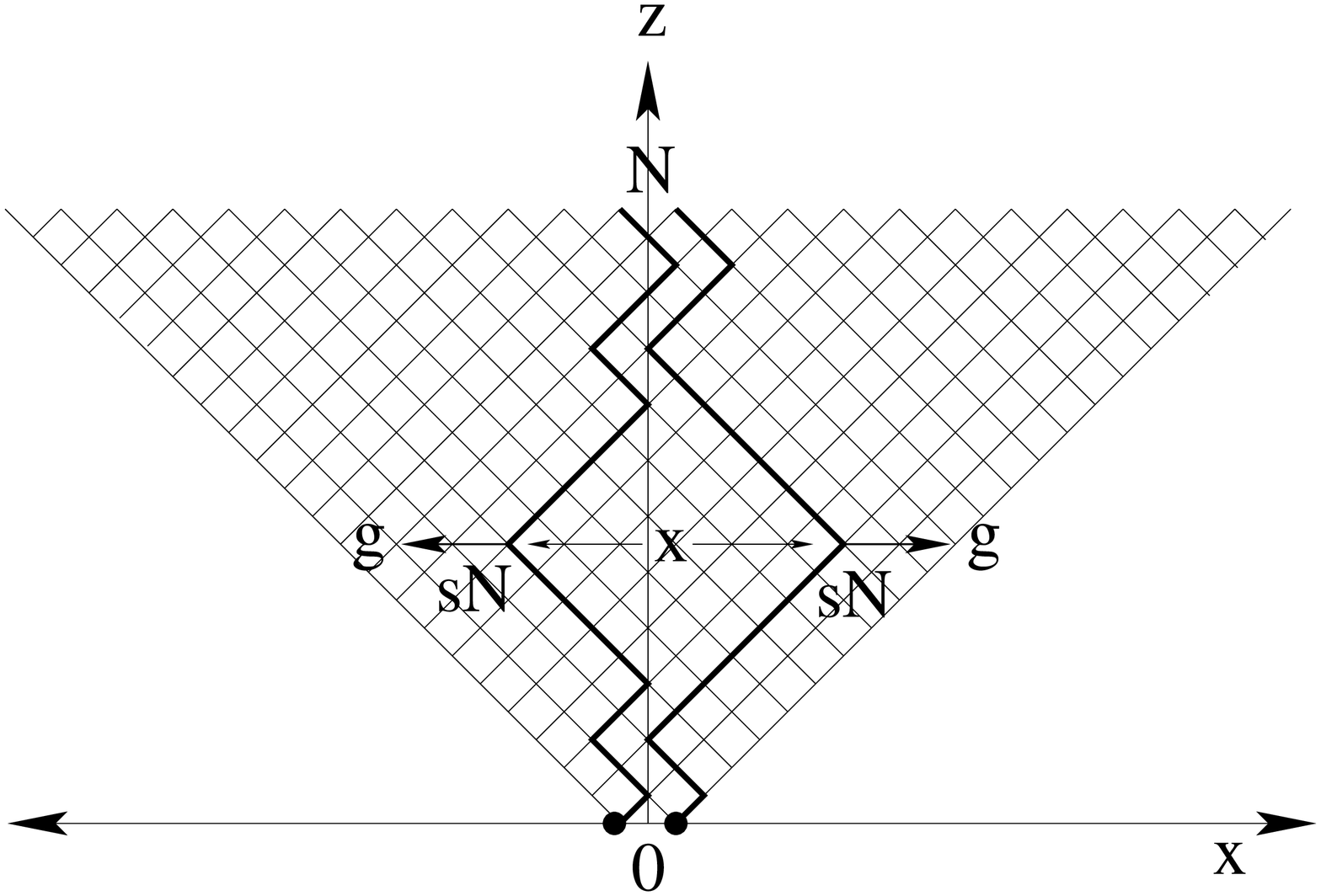}
    \caption{ \label{fig:schdia} Schematic diagram of DNA unzipping by a
	pulling force at a fraction $s$ ($0 \le s \le 1$) from the anchored
	end. In the fixed force  ensemble the force $g$ is kept fixed while
	the separation $x$ is  kept fixed in the fixed distance ensemble. }
\end{center}
\end{figure}
}

\newcommand{\figphasetwo}{
\begin{figure}[t]
\begin{center}
	\includegraphics[width=6.1cm]{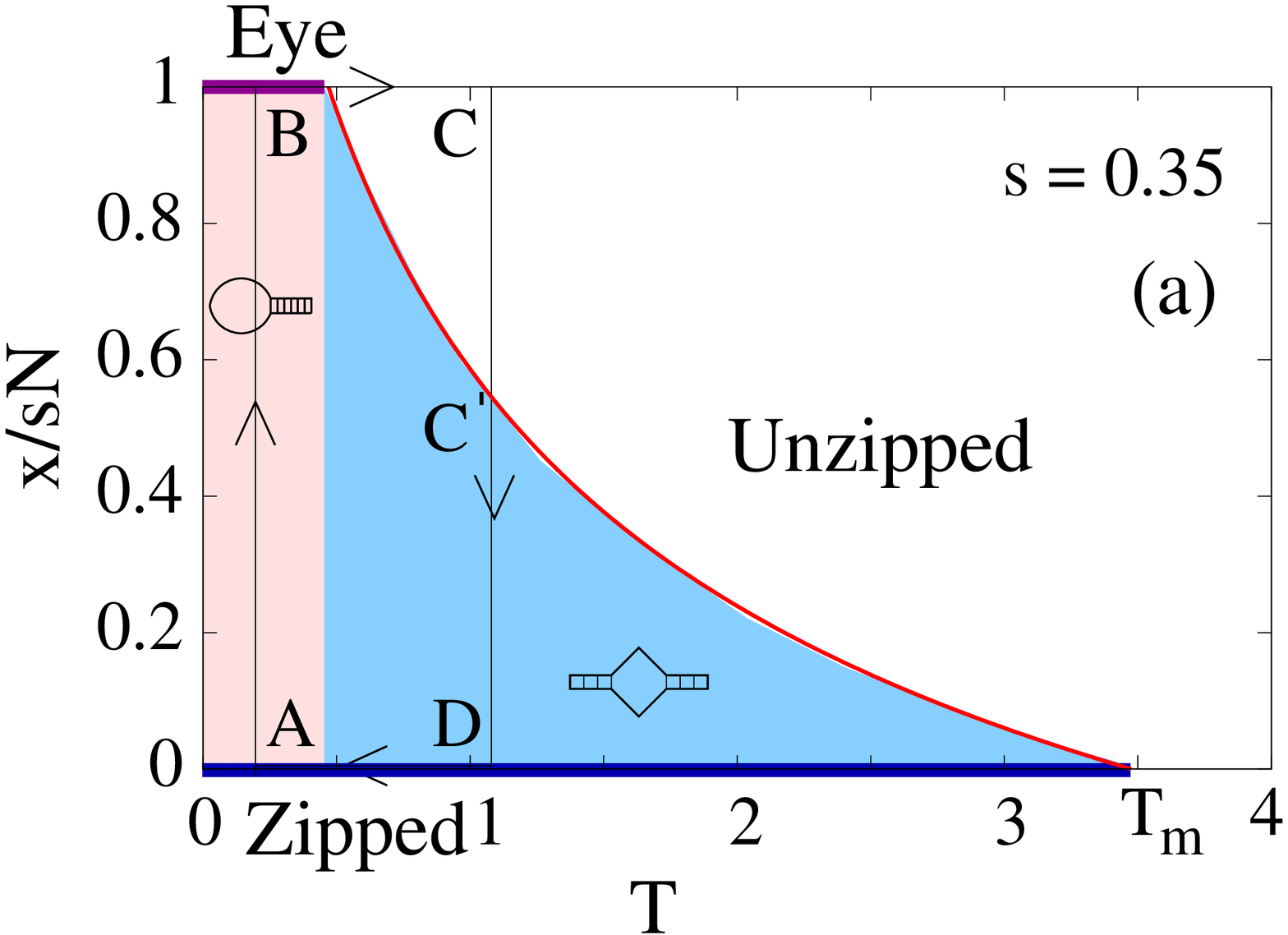}
    \includegraphics[width=6cm]{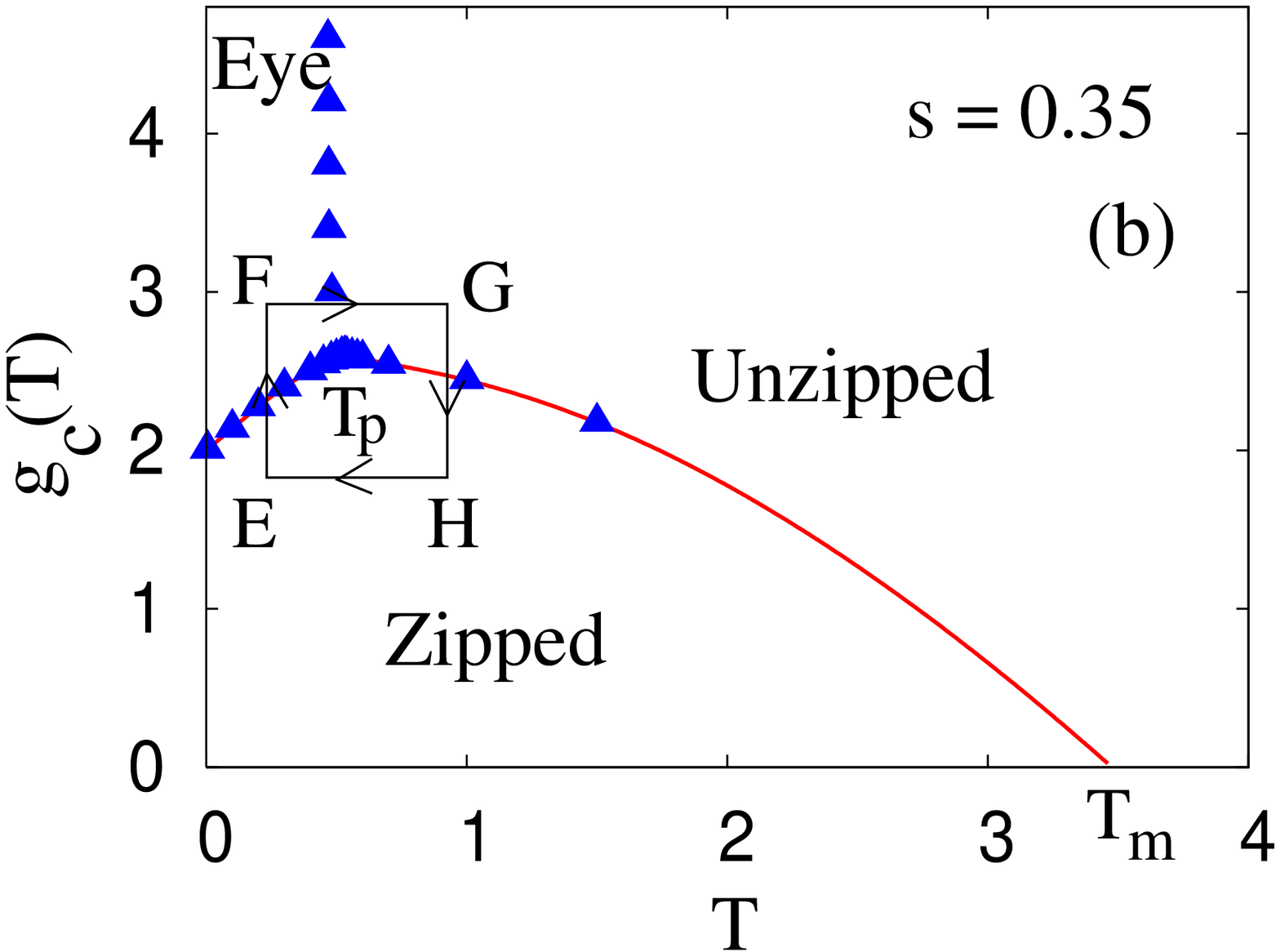}
	\caption{ \label{fig:phasetwo} The phase diagrams for $s=0.35$. (a)
	Fixed distance ensemble and (b) fixed force ensemble.  }
\end{center}
\end{figure}
}

\newcommand{\figphaseone}{
\begin{figure}[t]
\begin{center}
    \includegraphics[width=6cm]{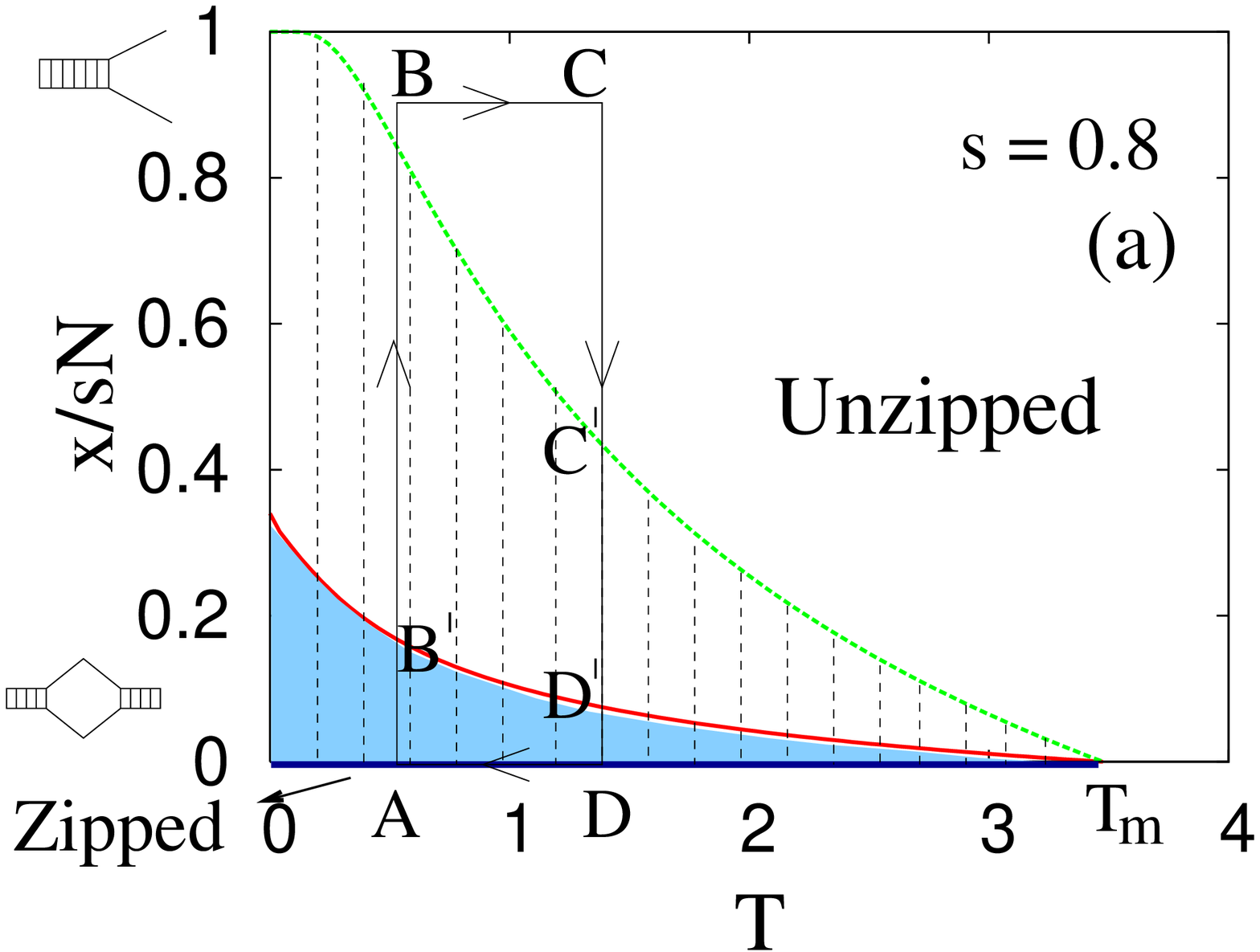}
    \includegraphics[width=6cm]{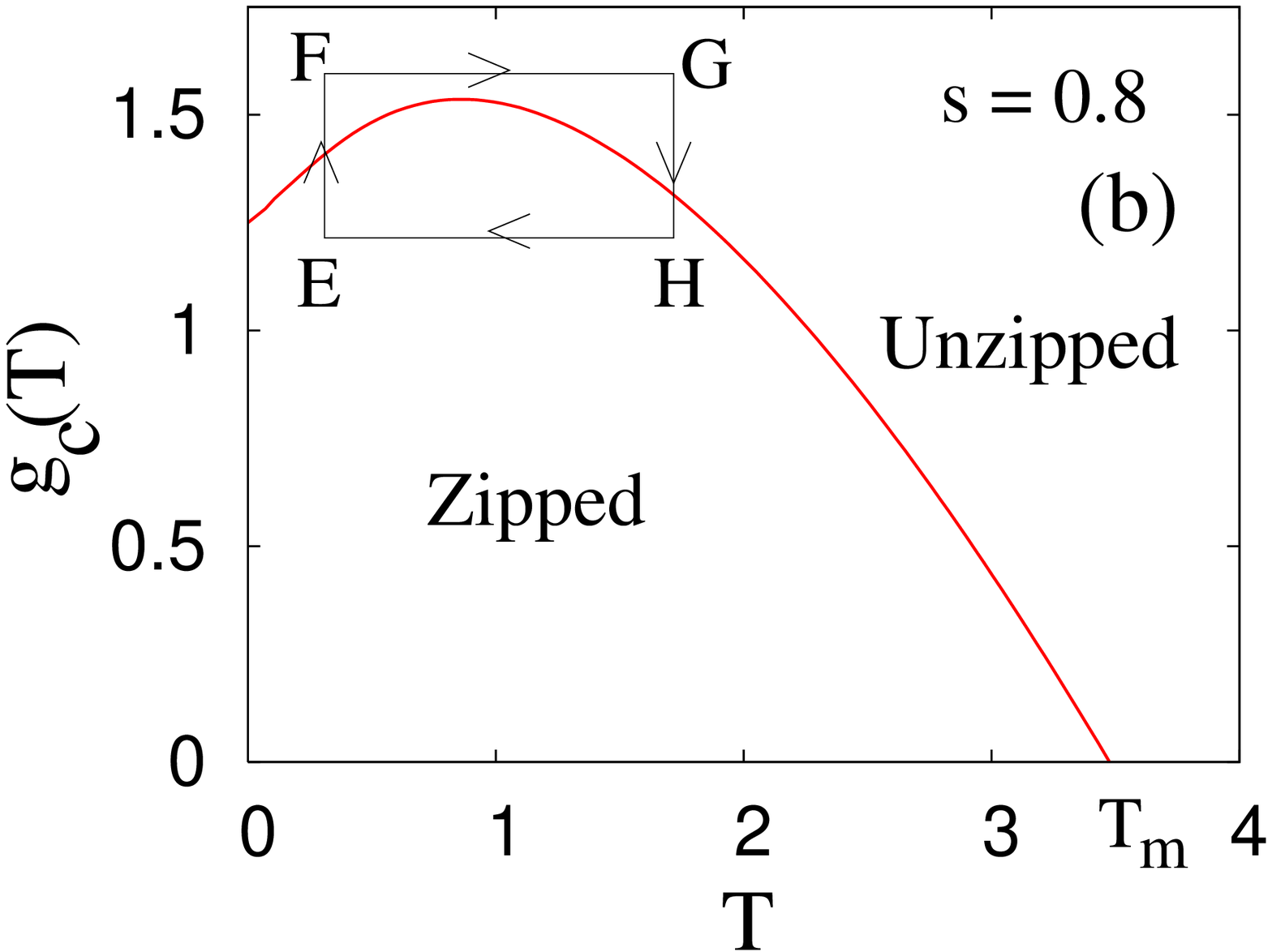}
\caption{ \label{fig:phaseone} The phase diagrams for $s=0.8$. (a) Fixed
distance ensemble and (b) fixed force ensemble.  }
\end{center}
\end{figure}
}

\newcommand{\figphds}{
\begin{figure}[t]
\begin{center}
   \includegraphics[width=6cm]{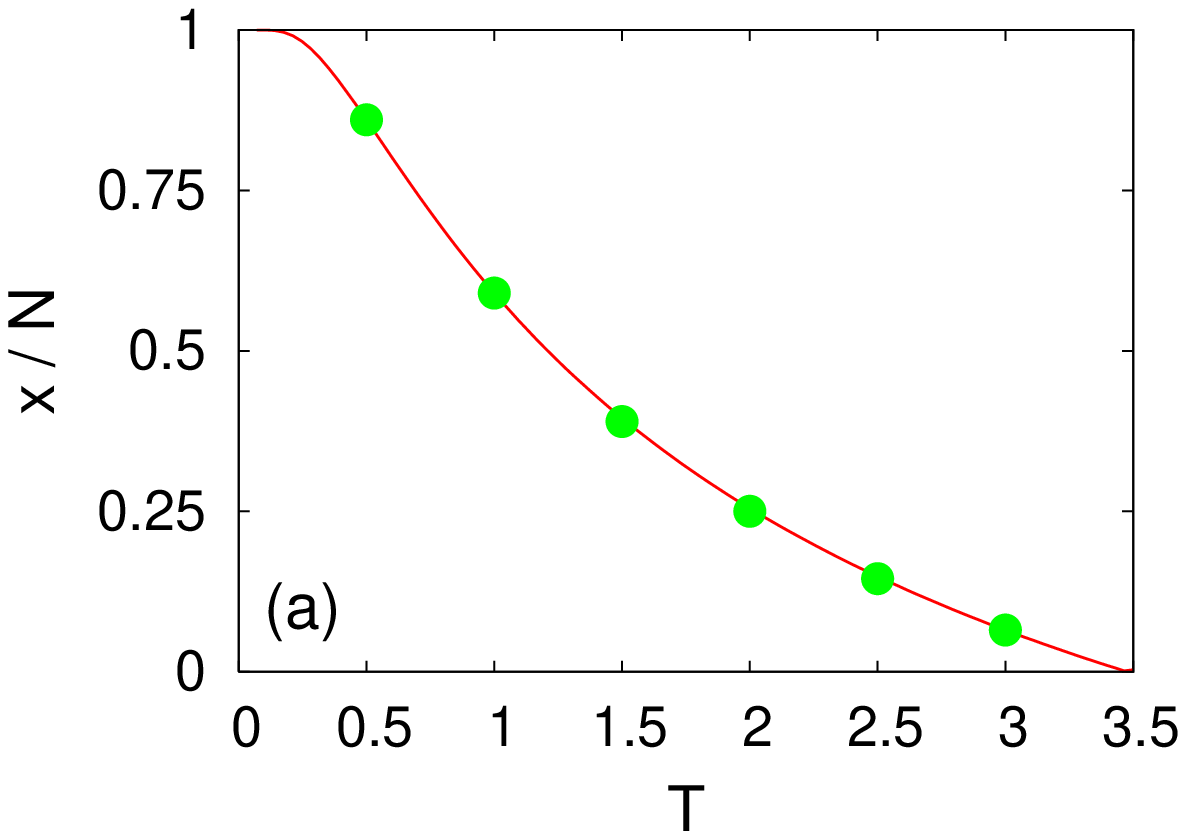}
   \includegraphics[width=6cm]{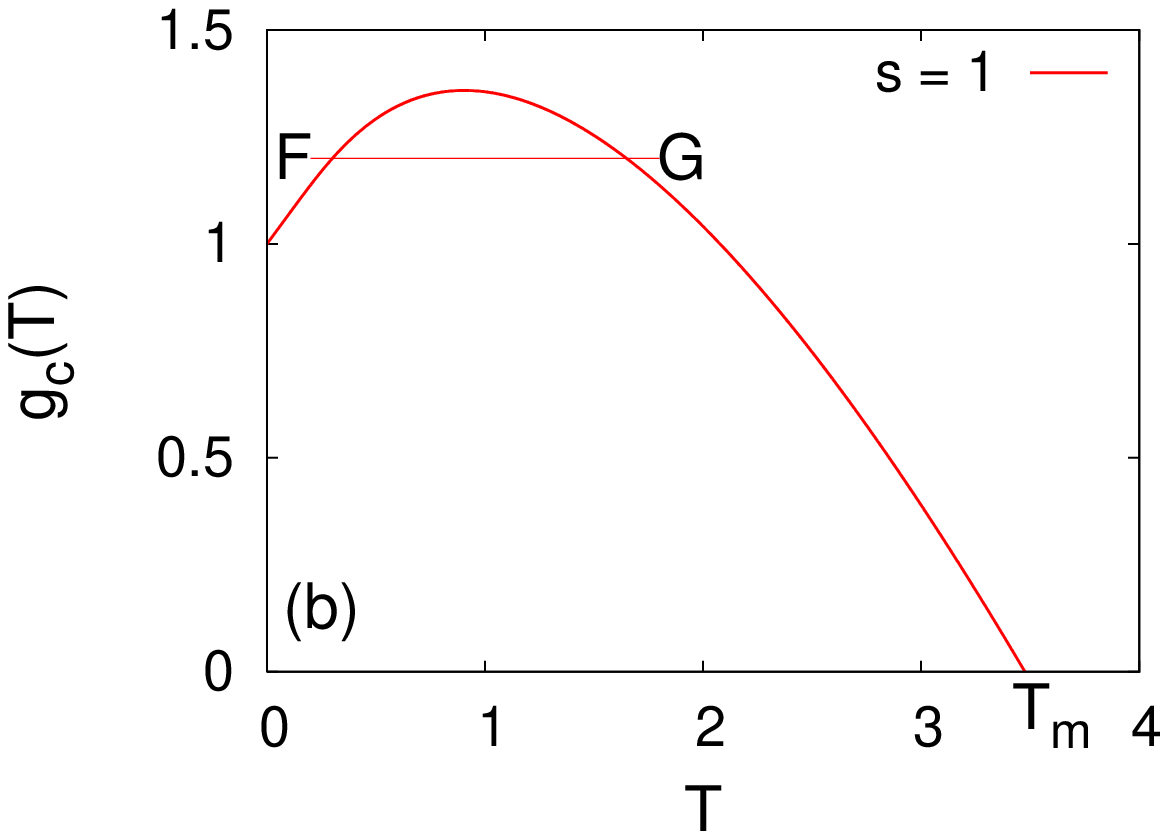}
   \caption{ \label{fig:seq11} (a) The coexistance region between the
   zipped and the unzipped phases when the separation between the last
   monomers ($s=1$) of the DNA is kept constant.  The points are from the
   transfer matrix based  numerical approach as discussed in the text and
   the solid line is the analytical result. (b) Phase diagram in the
   $g$--$T$ plane in the fixed force ensemble. The path FG is used for
   specific heat.  }
\end{center}
\end{figure}
}%

\newcommand{\figslope}{
\begin{figure}[t]
\begin{center}
    \includegraphics[width=4.0cm]{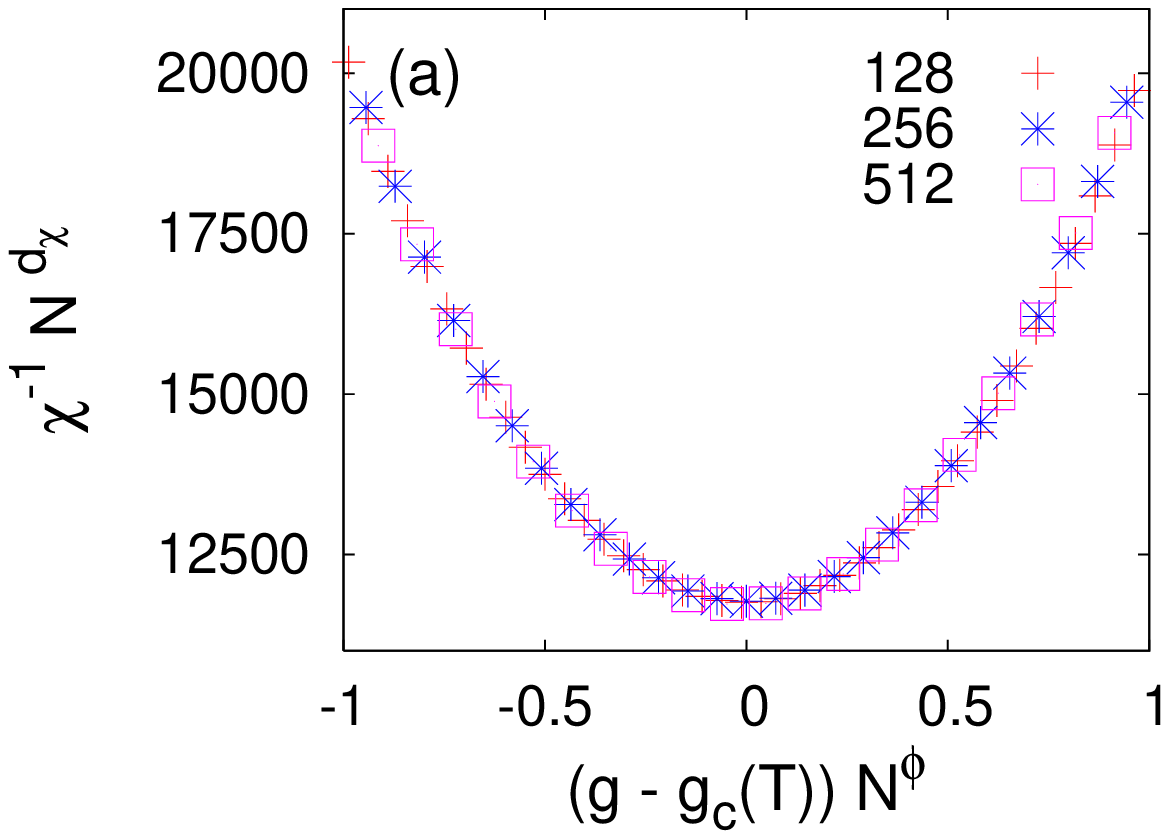}
    \includegraphics[width=4.0cm]{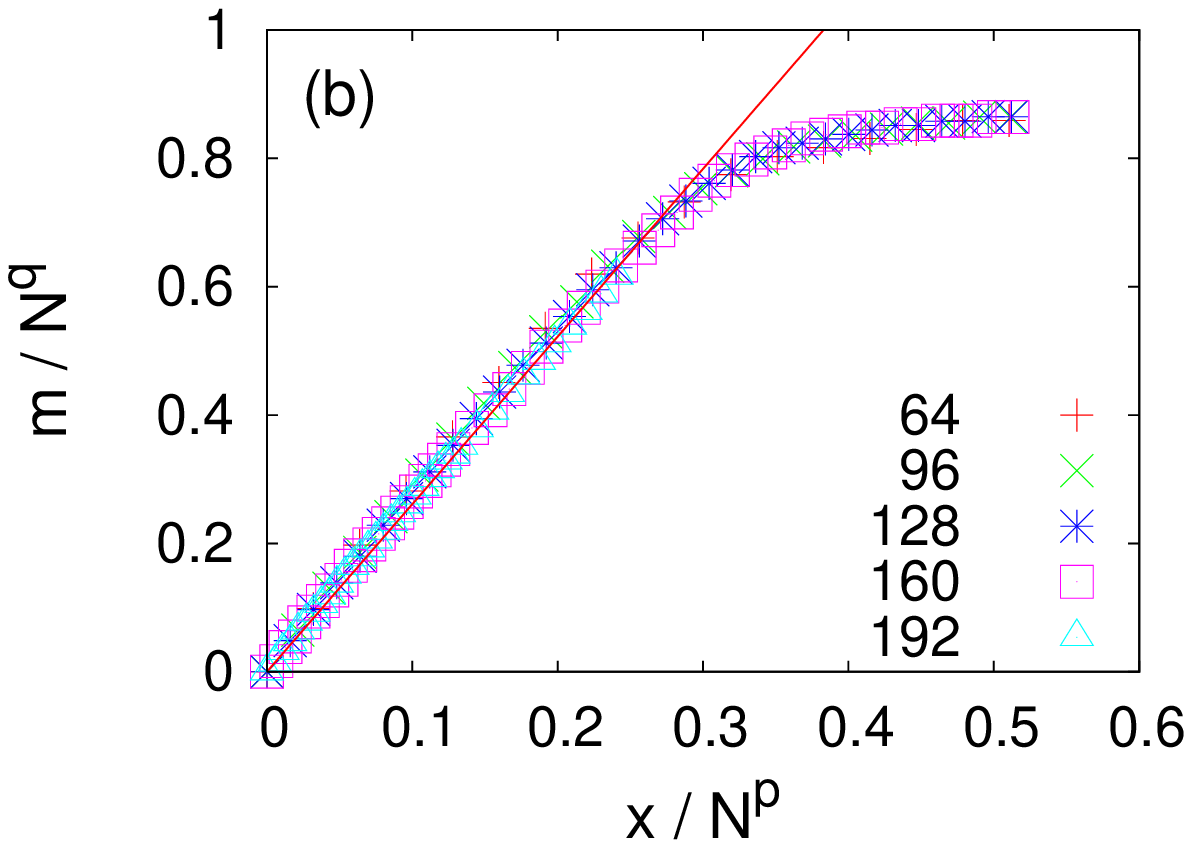}
    \includegraphics[width=4.0cm]{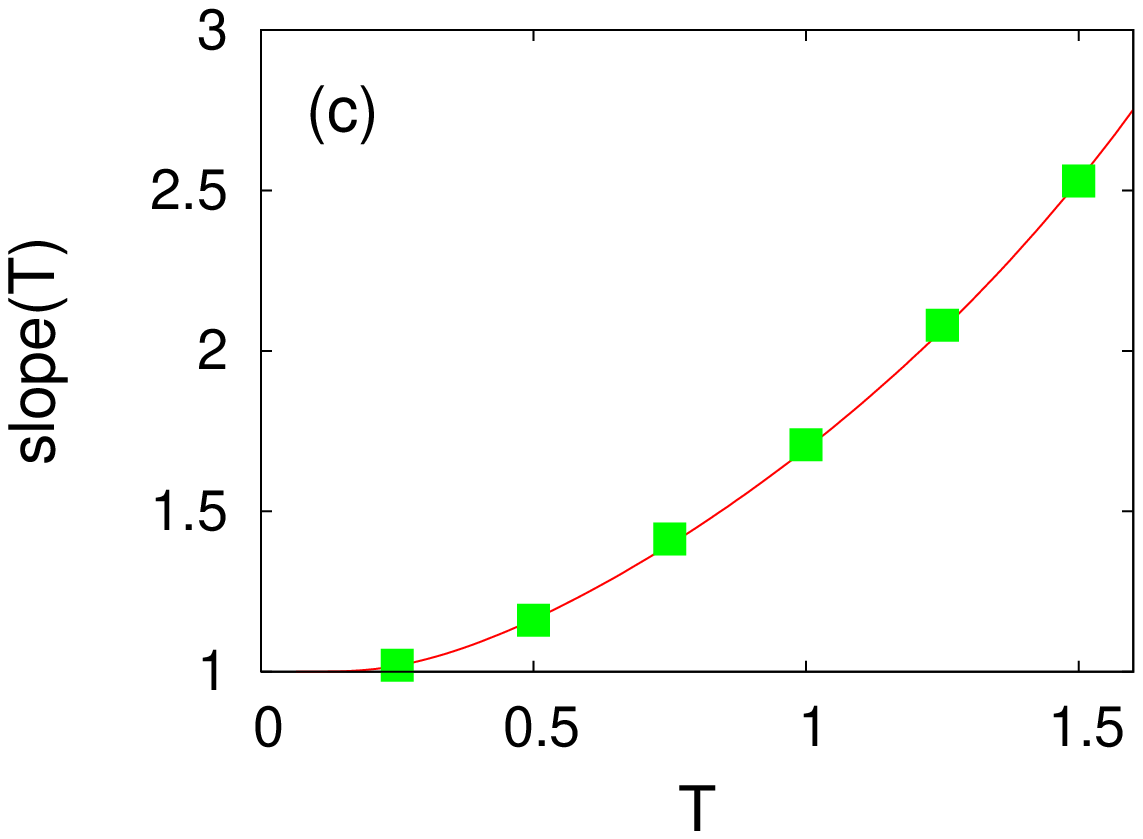}
	\caption{\label{fig:slope} (a) Data collapse of the inverse
	extensibility at $T=0.5$ for various lengths and forces.
	(b) Data collapse at $T=1.5$ for length $m$ of  Y-fork with
	separation $x$ between the end monomers of two  strands.  The
	straight line is the best fit to the linear region of the collapse
	and different symbols represent different 	chain  lengths.
	(c) The slope of the linear region of the collapse as a function of
	temperature. The points are from numerics and the solid line is the
	analytical result.  }
\end{center}
\end{figure}
}

\newcommand{\figspht}{
\begin{figure}[t]
\begin{center}
    \includegraphics[width=6cm]{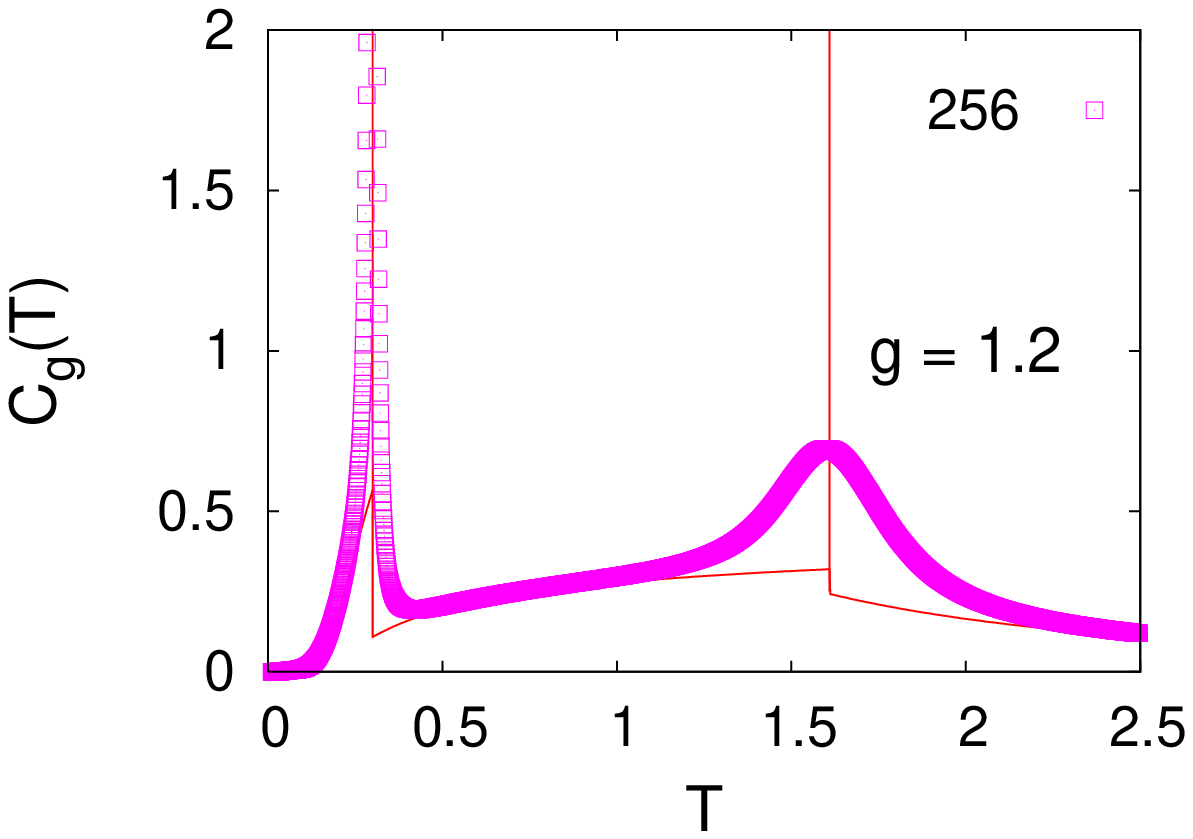}
	\caption{ \label{fig:spht}Specific heat $c_g$ as a function of
	temperature for a value of $g$ that allows reentrance (path FG in
	Fig. \ref{fig:seq11}). Solid line is from analytical results
        in the $N\to\infty$ limit  and points are from numerical results
		for a finite chain of length $N=256$. }
\end{center}
\end{figure}
}

\begin{document}

\paper[Unzipping DNA]{Unzipping DNA by force: thermodynamics and finite size
  behaviour}

  \author{ Rajeev Kapri$^{1}$ and  Somendra M.~Bhattacharjee$^{1,2}$ }
  \address{1. Institute of Physics, Bhubaneswar 751 005, India.\\
  2. TCMP, Saha Institute of Nuclear Physics, 1/AF Bidhannagar,
  Kolkata 700 064, India.\\
  email: rajeev@iopb.res.in, somen@iopb.res.in }

\begin{abstract}
  We discuss the thermodynamic behaviour near the force induced
  unzipping transition of a double stranded DNA in two different
  ensembles.  The Y-fork is identified as the coexisting phases in the
  fixed distance ensemble.  From finite size scaling of
  thermodynamic quantities like the extensibility, the length of the
  unzipped segment of a Y-fork, the phase diagram can be recovered.
  We suggest that such procedures could be used to obtain the
  thermodynamic phase diagram from experiments on finite length DNA.
\end{abstract}
\date{\today} 
\submitto{\JPCM}

\maketitle

\section{Introduction}
The force induced unzipping transition\cite{somen:unzip} provides a
mechanism to separate the two strands of a double stranded DNA without
a drastic change of environment like changing the temperature or pH.
In the last few years, the transition has been studied theoretically
[1-15]
and experimentally \cite{essevaz:micro,danil}.  The motion of a helicase
has also been studied with phase coexistence at unzipping as the
underlying mechanism\cite{helicase1,helicase2}.

With the advent of single molecule experiments, it is now possible to
explore the behaviour of a DNA be it single or with other
complexes\cite{essevaz:micro,kowal,lohman}.  These experiments, and
also biological molecules, work effectively in different ensembles.
For example atomic force microscope (AFM) works in the fixed distance ensemble
while several magnetic bead methods use the fixed force ensemble.
Helicases like dnaB work in the fixed distance ensemble, but PcrA type
helicases have both the fixed-distance and the fixed-force character (see,
e.g., the references in Ref. \cite{helicase1,helicase2}).
Consequently, these explore or take advantage of the various features
of the phase diagrams.  With this in mind, we like to discuss the
thermodynamic behaviour near the unzipping transition under either an
applied fixed pulling force (fixed force ensemble) or a fixed distance
between the strands (fixed distance ensemble) at a preassigned
fraction $s$ from the anchored end. After a brief review of the known
phase diagrams in Secs 2 and 3, we show how the results from finite
length DNA can be used to extract the phase diagram and also to determine
other thermodynamic quantities.  An interesting, and surely
realizable, situation would be a closed cycle in the thermodynamic
plane with the two ensembles revealing two types of features.  These
features are verifiable in experiments.

\figsch

\section{Model and ensembles}

The two strands of a homo-polymer DNA are mimicked by two directed
random walks on a $d=2$ dimensional square lattice. The walks start
from the origin and are allowed to go towards the positive direction
of the diagonal axis (z-direction) {\it without crossing each other}.
The non-crossing constraint is important for real DNA. The base
pairing between the two strands of DNA is modeled by a contact energy
$-\epsilon$ ($\epsilon >0$) for every contact (i.e. separation $x=1$,
see Fig. \ref{fig:schdia}).
The directional nature of the walks takes care of the correct base
pairing of DNA. In addition to this bonding, either a constant force
$g$ acts along the transverse direction ($x$-direction) or the
separation between the strands $x$ is kept constant at a fixed
fraction $s$ ($0 \leq s \leq 1$) from the anchored end ($z=0$). The
schematic diagram of our model is shown in Fig. \ref{fig:schdia}.

The fixed distance and the fixed force ensembles at temperature $T$
and chain length $N$, are characterized by the partition functions
\begin{equation}
\label{eq:10}
 Z_N(x,T)=\sum \exp(n\beta\epsilon),\ {\rm and}\ {\cal Z}_N(g,T)=\sum_x Z_N(x,T) \exp(\beta g x), 
\end{equation}
respectively, where, in $Z_N$, the sum is over only those
configurations which have the separation $x$ at the point in question
(fixed distance), while in ${\cal Z}_N$, the sum is over all
configurations with $n$ base-pairs and separation $x$ with summations
over all allowed values of $n$ and $x$, for a given force $g$ (fixed
force). Here $\beta=1/k_{\rm B}T$ is the inverse temperature, $k_{\rm
  B}$ being the Boltzmann constant.  We may choose a unit with $k_{\rm
  B}=1$ and $\epsilon=1$.  The free energies are given by
\begin{equation}
    \label{eq:14}
 F_N(x,T)=-k_{\rm B}T \ln Z_N(x,T)\ {\rm and}\    {\cal F}_N(g,T)=-k_{\rm B}T \ \ln {\cal Z}_N(g,T),
\end{equation}
respectively.

In the fixed force ensemble, the separation fluctuates, and one
defines the average separation $\langle x\rangle$, but in the fixed distance ensemble, one needs the average force to maintain the separation.  These in the corresponding ensembles are 
given by
\begin{equation}
  \label{eq:13}
  \langle x \rangle = -\frac{\partial {\cal F}_N(g,T)}{\partial g}, \ {\rm and}\ 
  \langle g\rangle=\frac{\partial F_N(x,T)}{\partial x}.
\end{equation}
at a fixed $T$, with the angular brackets $\langle ...\rangle$
denoting thermal averaging in the appropriate ensemble.

In all these cases, the phase transition is obtained only in the limit
$N\to\infty$.  We show how finite size scaling can be used to extract
the $N\to\infty$ phase diagram and other information from finite $N$
results.  Since experiments are done on short DNA, we suggest
that a similar approach should be adopted to infer the thermodynamic
behaviour.

\section{Phase Diagrams}
Let us recapitulate the phase diagrams for this model for a few
typical values of $s$ in two different ensembles\cite{scaling,kapri}.

\subsection{$s=1$: pulling at the end}
\label{sec:seq1}

\figphds

In the case where the force is applied at the free end, there are only
two possible phases, namely the zipped and the unzipped phases.  The
free energies per base-pair of these two phases are
given by
\begin{equation}
  \label{eq:1}
  f_{\rm zp}(T,s=1)=-k_{\rm B}T \ln z_3(T), \ {\rm and} \   f_{\rm
    uz}(T,s=1)=-k_{\rm B}T \ln z_2(T),
\end{equation}
where
\begin{equation} 
\label{eq:z3}
z_3(T)=\sqrt{1-e^{-\beta \epsilon}} -
1 + e^{-\beta \epsilon} \ {\rm and}\   z_2=(2+2\cosh{\beta g})^{-1}.
\end{equation}
The force-temperature phase boundary between the two phases in the
fixed force ensemble is given by
\begin{eqnarray}
  \label{eq:2}
  g_c(T) &=& k_{\rm B} T \cosh^{-1} \left [ \frac{1}{2} \frac{1}{ \sqrt{-e^{-\beta
  \epsilon } +1} - 1 + e^{-\beta \epsilon }} -1 \right ] \nonumber \\
  &=& - k_{\rm B} T \ln \lambda(z_3),
\end{eqnarray}
where $\lambda(z) = (1-2z -\sqrt{1-4z})/(2z)$.  This boundary is
shown in Fig. \ref{fig:seq11}(b).

In the absence of any force, the DNA can be denatured (melting) simply
by increasing the temperature. The denaturation (melting) temperature,
$T_m=1/\ln(4/3)=3.476059497...$ corresponds to a pure second-order
phase transition with a discontinuity in the specific heat. {\it The
  unzipping transition is however first order}. The end point
separation of the two strands on the phase boundary jumps from a small
value in the zipped phase to an extensive number ({\it i.e.},
proportional to the length)
\begin{equation}
  \label{eq:3}
  X_c(T)\equiv x/N= \tanh [g_c(T)/T].
\end{equation}

In the fixed distance ensemble, the phase diagram shows the
coexistence region demarcated by the line $X_c(T)$ of Eq.(\ref{eq:3}),
as shown in Fig. \ref{fig:seq11}(a). {\it This coexistence region
  resembles the Y-fork structure} formed during DNA
replication\cite{alberts03}.

\subsection{$s >1/2$}
\label{sec:s-12}

Figure \ref{fig:phaseone} shows the phase diagrams for $s=0.8$ in both
the fixed distance and the fixed force ensemble. These are representatives
of all $s >1/2$, for which the double stranded DNA can be fully
unzipped at low temperatures either by increasing the separation
between the two strands in the fixed distance ensemble or by increasing
the force in the fixed force ensemble. In the fixed force ensemble,
the DNA is either in the zipped or in the unzipped phase depending on
the applied force and the working temperature.  The free energy per base
pair for the unzipped phase is now
\begin{equation}
  \label{eq:4}
   f_{\rm uz}(T,s)=-k_{\rm B}T \ln z_2(T,s), \quad z_2(T,s)=[4^{1-s}(2+2\cosh{\beta g})^s]^{-1}.  
\end{equation}
The $s$-dependent phase boundary is given by
\begin{equation}
\label{unzippb}
g_c(s,T)= k_{\rm B} T \ln \lambda(4^{(1-s)/s} z_3(T)^{1/s}).
\end{equation}
The difference (with Eq. \ref{eq:2})  owes its origin to the extra entropy of the unzipped
chain between the point of application of the force and the free end.

\figphaseone

However an eye of the type of Fig. \ref{fig:schdia} is possible in the
fixed distance ensemble. This is one of the ensemble differences.  The
phase diagram in the fixed distance ensemble shows two possible
coexistence regions : (i) A region (shown by vertical lines) in
which one end of the DNA is unzipped into two single strands and other is
a double strand. Such configurations resemble the Y-fork as in the
$s=1$ case. (ii) A region (shown by shade) having the configurations
of the type shown in Fig.  \ref{fig:schdia}. These configurations
resemble the transcription bubble formed, e.g. by RNA polymerase
during RNA transcription.

Quantitatively,  for
\begin{equation}
  \label{step}
  \frac{x}{sN} < X(s,T)\equiv \frac{1-s}{s} \frac{\ln (4z_3)}{\ln
  (\lambda(z_3))},
\end{equation}
one finds bubbles beyond which the free end unzips yielding the Y-type
configurations, though with free wings.  In terms of the force
required to maintain the distance, there will be two situations.
\begin{eqnarray}
\label{pbs}
        g_c(s,T)&=&2g_c(T), \ {\rm if\ Eq.(\ref{step})\   is\  satisfied}\\
        \label{pbsb}
        &=& g_c(T), \ {\rm otherwise}.
\end{eqnarray}
With the increase of the separation $x$, the end point gets detached
at the critical value $x=sNX(s,T)$, provided $X(s,T)<1$.  Once all the
bonds are broken the two open tails behave like free independent
chains.  In such a situation, the force required to maintain the
separation is just like the $s=1$ end case (in the $sN \rightarrow
\infty$ limit) in Eq.(\ref{pbs}).  For this to happen we also require
$x/sN < X_c(T)$, else we have the unzipped phase.

\subsection{$s<1/2$}
\label{sec:s12}

If a force is applied at $s<1/2$, a new phase occurs, namely the
``eye'' phase which involves the anchored point at $z=0$ as one of its
extremities. The phase diagrams in the two ensembles are shown in
Fig. \ref{fig:phasetwo}.  The boundary separating the zipped from the
unzipped phase 
matches with the boundary we already derived in Eq.(\ref{pbs}).  This
transition to the eye phase in presence of a force is also found in
the fixed distance ensemble. The zipped-unzipped phase boundary
remains the same as Eq.(\ref{unzippb}). The bubbles are formed in the
zipped-unzipped transition.  Because of the presence of a third phase, a
triple point appears in the phase diagram, easily identifiable in fixed
force ensemble phase diagram for $s=0.35$ (Fig. \ref{fig:phasetwo}(b)).

\figphasetwo

\section{Finite size scaling}
The response of the DNA to the pulling force can be defined by the
extensibility $\chi$ as
\begin{equation}
  \label{eq:5}
  \chi_T=\left .\frac{\partial x}{\partial g}\right|_T,
\end{equation}
which in the fixed force ensemble can be related to the fluctuation in $x$,
\begin{equation}
  \label{eq:6}
  \chi_T=\frac{1}{k_{\rm B}T} (\langle x^2\rangle - \langle x\rangle^2).
\end{equation}
This extensibility is an extensive quantity, proportional to the
length of the chain.  For convenience, we shall omit the subscript $T$
unless required.

A phase coexistence is characterized by a flat $g-x$ isotherm.  This
means a finite change in $x$ can be induced without change in $g$
yielding an infinite $\chi$. A diverging $\chi$ is however not
possible for finite chains.  Instead a finite-size scaling
behaviour is expected.  Such a scaling form would allow one to
determine the phase boundary from finite size data and would be useful
in cases where exact solutions are not known.


The free energy at a given value of $x$ (with say $s=1$) for a finite
value of $N$ can be determined with arbitrary precision by a transfer
matrix approach.  By taking finite differences of the free energy for
different values of $x$ we can determine the force and the
extensibility at a given temperature.  The inverse extensibility shows
the finite size scaling form
\begin{equation} 
\chi^{-1} =N^{-d_{\chi}} {\cal{F}}\left( (g -g_c(T))N^{\phi} \right),
\end{equation}
and a data collapse is obtained for $d_{\chi}=2$ and $\phi=1$ as
determined by the Bhattacharjee-Seno method\cite{collap}.  The data
collapse is shown in Fig. \ref{fig:slope}(a).  Keeping in mind that $\chi$ is
extensive, these exponents show that the extensibility has a
divergence as
\begin{equation}
  \label{eq:7}
  \chi\sim \mid g-g_c(T)\mid^{-1},
\end{equation}
as has been predicted in continuum models\cite{somen:unzip}.

A similar scaling form with $x$ as the variable allows us to locate
the end point of the coexistence region in the fixed distance
ensemble.  We used the same procedure at different temperatures and
trace out the coexistence region for the DNA problem in fixed distance
ensemble. The results (circles) obtained via this procedure agree very
well with the exact results (solid curve) in Fig. \ref{fig:seq11}.
That the finite size exponent $\phi=1$ in Eq.(\ref{eq:7}) follows from
a careful analysis of the singularities of the generating function
used for the exact solution\cite{scaling,kapri}.

\figslope

Another quantity of interest is the length $m$, of the unzipped
segment in the Y-fork as a function of separation $x$ between the end
monomers of two strands of DNA at a fixed temperature.  A phase
coexistence means the DNA is segregated into an unzipped region of
length $m$ and a zipped region of length $N-m$. If the equation of
state of the unzipped chain of length $N$ be written as $x=N {\cal
  A}({g/T})$, then for a given separation $x$, one expects the length
of the unzipped chain to be
\begin{equation}
  \label{eq:8}
    m(x)= x/{\cal A}(g_c(T)/T)=x/\tanh (g_c(T)/T),
\end{equation}
where we have used the equation of state as given by Eq.(\ref{eq:3}).


From the transfer matrix approach, the length of the unzipped chain of
the Y-fork can be determined.  These exhibit the finite size scaling
of the form
\begin{equation}
m(x) = N^{q}{\cal{M}}(x/N^p)
\end{equation}
with $q=p=1$. Figure \ref{fig:slope}(b) shows the data collapse at
$T=1.5$ for different chain lengths.  We find that $m(x)$ behaves
linearly for small end separation $x$. The straight line is the best
fit to this linear region. In Fig. \ref{fig:slope}(c) we have plotted
the slope of this linear region as a function of temperature. The
points are the slope of the fitted line and the solid curve is the
known analytical result of Eq.(\ref{eq:8}).  This result corroborates
the basic tenet of {\it coexistence} of the two phases on the one
dimensional DNA chain.

\subsection{Specific heat}
The divergence of $\chi$ and the temperature dependence of $m$ at
coexistence can be used to determine the specific heat behaviour.
There are two specific heats to be considered, the constant $x$
specific heat $c_x$ and the constant force specific heat $c_g$.

The constant $x$ specific heat (per base-pair) follows from the bulk
free energy
\begin{equation}
  \label{eq:9}
  Nf(T,x)= m f_{\rm uz}(T) + (N-m)f_{\rm zp}(T,x).
\end{equation}
By taking derivatives and noting that the two free energies are same at
coexistence, the constant $x$ specific heat $c_x(T)$ close to the
transition is given by
\begin{equation}
  \label{eq:11}
  c_x(T)\approx \frac{m}{N} c_{x,{\rm uz}}(T) + \frac{N-m}{N}c_{\rm zp}(T) + 
  \frac{dm}{dT} {\cal L},
\end{equation}
where ${\cal L}$ is the latent heat.  At constant $x$, as the
temperature is changed, the heat supplied goes partly in changing the
temperature of the individual components and also in converting a part
of the zipped chain to unzipped, i.e. to change $m$. For the last
process, a latent heat is required.  This is the content of the above
equation Eq.(\ref{eq:11}).  As the phase boundary is approached, $m/N$
approaches $1$ smoothly, see Eq.(\ref{eq:8}). Beyond the phase
boundary, it is only the unzipped chain.  Therefore, the constant $x$
specific heat as a function of $T$ shows a {\it jump discontinuity} at
the phase transition point though it is a first order transition. The
peculiarity comes from the fact that for the most of the temperature
range in the constant distance ensemble, we see phase coexistence,
i.e. a Y-fork.

If we consider the specific heat at a constant force, then we cross
the phase boundary at the transition point only while at all other
temperatures we are strictly in one phase. Therefore, below the
transition point ($T<T_c(g)$) we see the specific heat of the zipped
chain while above ($T>T_c(g)$) that of the unzipped chain. At the
transition point for the given force ($T=T_c(g)$), the latent heat
supplied goes in changing the phase but not the temperature.  Consequently, 
a delta function peak (divergence) appears in $c_g$ at $T=T_c(g)$.  This
divergence of $c_g$ also follows from the thermodynamic relation of
the two specific heats,
\begin{equation}
  \label{eq:12}
  c_g-c_x = T \chi_T \left(\left.\frac{\partial g_c(T)}
          {\partial T}\right|_{x}\right )^2.
\end{equation}
Since $c_x$ is finite at all $T$, we see the divergence of $c_g$ at
$T_c(g)$ is linked to the divergence of $\chi$ at that point.  A
similar analysis can be done for the reentrant part of the phase
boundary also.

The analytical results for the thermodynamic limit and the numerically
calculated specific heat for a finite length of DNA with a force that
admits reentrance is shown in Fig. \ref{fig:spht}.  The two peaks for
the two transitions are prominent in the finite size results.

\figspht

\section{Closed loops in the phase diagram} 
The $s$-dependent phase diagrams can be used to interpret the effect of
helicases and other motor proteins on DNA which keep the separation
between the two strands constant.  They may be classified as
fixed-distance objects (FDO's).  Take for example dnaA which is known
to initiate the replication process.  It generally starts at a point
near an end which means $s<1$. As it starts moving towards the end, $s$
starts increasing and the $s$-dependence of the phase boundary of Fig.
\ref{fig:phaseone} (Eq.(\ref{unzippb})) shows that at some value of
$s$ for the given $T$, the end points get released. This is accompanied
by a  sudden change in the force (Eqs.(\ref{pbs}),(\ref{pbsb})) on the
FDO and this sudden change  can be 
responsible for its offloading. Similarly, helicases have components to
maintain a fixed distance between the two strands to behave as FDO's.
The motor action and any further destabilizing effects (like force in
the case of pcrA) is equivalent to a thermodynamic force on the junction
of the two coexisting phases at the Y-fork (``domain wall'').  The
opening of the DNA by a helicase can then be interpreted as a motion of
the wall or a propagating front through the double stranded DNA under a
local time-dependent instability. The velocity of the front (i.e.
opening of the DNA) can be related to the elastic behaviour of the
wall\cite{helicase2}.

Let us now discuss the features that one should see if one moves in a
closed loop on the phase diagram.

Suppose we start with a DNA in its zipped phase (point A in Fig.
\ref{fig:phaseone}(a)) in the fixed distance ensemble and use a tip (say
of AFM) to increase isothermally the separation between the monomers
of two strands at a fraction $s>1/2$ from the anchored end.  As we
move on line AB in the phase diagram, a bubble starts emerging for
small $x$.  The size (i.e. length) of the bubble grows with separation
until we cross the solid curve at B$^\prime$ where the free end of the
DNA gets unzipped into two single strands. This curve (for B$^\prime$)
represents the jump from one coexistence region to another with a
{\it sudden, detectable  change in force} (Eqs.(\ref{pbs}) and
(\ref{pbsb})).  Point B$^\prime$ corresponds to the separation $X(s,T)$
of Eq.(\ref{step}). Once in the unzipped region at point B, we raise the
temperature to point C.  There is no change in the phase though the
elastic response of the unzipped chains will show a $T$-dependence. On
releasing the tip, i.e. along the path CD, we first see the formation of
a Y-fork at C$^\prime$ via zipping of the base-pairs {\it from the
anchored end} and a sudden change in force will be felt.  This force
remains constant until we reach the point D$^{\prime}$ when the free end
of the DNA zips leading to the formation of a bubble.  The force gets
doubled at this point. By a reduction of $T$ we may then close the loop
to get back the original state.

A closed loop EFGH of Fig. \ref{fig:phaseone}(b) will show only the
zipped and the unzipped phases, with the Y-fork appearing at special
values of the force and/or temperature.

A similar closed loop cycle may be performed for $ s < 1/2$.  Again we
start from the zipped phase at a low temperature (point A in Fig.
\ref{fig:phasetwo}(a) ) and
increase the separation between the strands at fraction $s<1/2$
keeping the temperature constant. For any $x$, we get the ``eye'' with
a zipped tail (``tadpole'').  This state is characterized by all open
bonds at the anchored end and the free end of DNA is still in its
zipped state. If the temperature is increased, the eye phase of DNA
gets denatured at a well defined temperature {\it which is below the
  bulk melting temperature $T_m$}.  At C one gets two single strands.
If we decrease $x$ at this new temperature (path CD) the two strands
of DNA start feeling each other's presence at C$^\prime$ where we
cross the solid curve at a (temperature dependent) critical separation
$X(s,T)$ at which both the ends of the DNA get re-zipped and a bubble
is formed on the DNA. By cooling one gets back into the eye phase.
Further decrease of separation brings back the DNA to its initial
zipped phase.

In the fixed force ensemble, we start from the zipped region at low
temperature (point E). The closed loop EFGH takes us from the zipped
to the eye (at F), then to the unzipped (at G) and then back to the
zipped phase (at H, E). We see a bubble formation only at specific
points on the cycle.

\section{Conclusion}
We have shown how the response function, extensibility, of a double
stranded DNA to a pulling force can be used to obtain the
thermodynamic phase diagram.  Our numerical procedure on short chains
reproduced the exact phase boundary.  We suggest that a similar
procedure be adopted for results from experiments which are
necessarily on short DNA chains.  The behaviour of specific heat is
also studied.  The signature of the phase diagram can be felt in a
thermodynamic cycle.  The differences in ensemble (fixed distance
versus fixed force) are also shown in various situations, especially
in closed loops.  These are verifiable in single molecule experiments
with DNA.

\end{document}